\documentstyle[aps,prb,twocolumn,psfig]{revtex}
\draft
\def\beq{\begin{equation}}
\def\eeq{\end{equation}}
\def\beqarr{\begin{eqnarray}}
\def\eeqarr{\end{eqnarray}}

\input epsf.tex

\begin{document}
\draft

\twocolumn[\hsize\textwidth\columnwidth\hsize\csname @twocolumnfalse\endcsname

\title{Deformation and Depinning of Superconducting Vortices from 
Artificial Defects:  A Ginzburg-Landau Study}
\author{D. J. Priour Jr$^{1,2}$ and H. A. Fertig$^1$}
\address{$^1$Department of Physics and Astronomy, University of Kentucky, Lexington, KY40506-0055\\
	 $^2$Center for Computational Sciences, University of Kentucky, Lexington, KY40506-0045}

\date{\today}

\maketitle

\begin{abstract}

Using Ginzburg-Landau
theory, we have performed detailed studies of   
vortices in the presence of artificial defect arrays,
for a thin film geometry.
We show that when a vortex approaches 
the vicinity of a defect, an abrupt transition occurs in which the vortex
core develops a ``string'' extending to the defect boundary, while
simultaneously the supercurrents and associated magnetic
flux spread out and engulf the defect.   
Current induced depinning of vortices is shown to be
dominated by the core string distortion in
typical experimental situations. 
Experimental consequences of this 
unusual depinning behavior are discussed.  

\end{abstract}


]

%
%
\section{Introduction}
One of the most important properties of superconductors
is their ability to carry currents without dissipation,
allowing them to generate large magnetic
fields.  Many superconductors allow fields to penetrate
in bundles of quantized magnetic flux, with associated whirlpools
of current known as vortices.  When these vortices are mobile, they
spoil the perfect conductivity that make superconductors so useful. 
The quest to increase the maximum dissipationless
current $J_{c}$ that a superconductor may carry has thus
fueled intense study of vortex pinning.

In recent years,   
pinning environments of artificially 
fabricated nanoscale defect arrays have been  
developed in hopes of better understanding and improving
the pinning properties of superconductors.  
Some of the earlier contributions have 
involved macroscopic measurements (e.g. $J_{c}$, magnetization) 
performed on periodic ``antidot'' arrays\cite{Harada,Baert2,MoshB96}. 
The antidot regions contain material which is
rendered non-superconducting. 
Pinning behavior may be studied in these periodic systems
using various imaging techniques\cite{Harada,Motton,Trapton,Field,Tona}. 
  

Much of the theoretical work on vortex pinning in defect arrays has 
employed numerical studies to focus\cite{reichhardt,crabtree} 
on the behavior of large collections
of vortices under the influence of a driving force
(supercurrent).  Molecular dynamics\cite{reichhardt,Nori} approaches
in particular usually
employ simplified pinning potentials, in part to 
make possible simulations of large numbers of
vortices, but also because information about
pinning potentials at the microscopic level
is simply unavailable.  A few studies\cite{buzdin,mkrtch} 
have focused on energy scales
for pinning varying numbers of flux quanta to
the defects, as well as defect-vortex potentials
as derived from the London equation\cite{mos,cheng}.
However, the latter approach does not allow for
variation of the Cooper pair density, and in particular
cannot correctly treat the vortex core.
In the London approach,  vortex cores are usually assumed to
be rigid in shape, and interactions of vortices with their environments are
determined by the core position as well as the distribution 
of currents\cite{mos,cheng}.
Our results demonstrate that the vortex core in fact deforms dramatically
near an artificial defect: when the vortex center is sufficiently close
to the defect, a {\it string} of suppressed order parameter
develops from the vortex position to the pinning center edge, while
the currents and magnetic flux spread out over a large area
(see Fig. 2 below).  

Usually, experimental studies of 
depinning are conducted in the presence of a net current across the 
superconducting film.  The
current exerts a Lorentz force on the flux quantum, dislodging  
the vortex from the defect for currents exceeding the depinning threshold.  
We find that, in 
addition to this, the external current distorts the conservative part of 
the pinning force 
for a sufficiently large net current, leading to   
important changes in the order parameter, supercurrents, and
depinning threshold.  
Specifically, we find for many typical experimental situations
that in the presence of a driving current, the core strings
associated with specific dots will stretch to reach 
neighboring dots.  At the depinning threshold, the current
associated with a vortex unwraps from one pinning site and 
encircles its neighbor, {\it without the order parameter 
ever forming a compact core region.}  Thus, for many experimental 
situations, interstitial vortices never form when vortices 
are depinned, as is commonly assumed.  The depinning process
should instead be understood as one in which the vortex 
cores are elongated by the driving current, eventually 
interacting with more than one pinning site at a time and 
allowing the associated vortex currents to hop from site
to site.

\section{Methods}

   Our calculations focus on two dimensional arrays
of artificial pinning centers in the form of
holes in a bulk superconductor.
Our goal is to find the lowest energy state
of the system for
specified locations of a superconducting vortex;
from this we can construct a pinning
potential.  The appropriate
description of the superconducting state is in terms
of Ginzburg-Landau theory, which focuses on
a complex superconducting order parameter $\psi(\vec{r})$,
for which $\left| \psi \right|^{2}$ is proportional to the 
local density of superconducting electrons.   Unlike the 
London theory, Ginzburg-Landau theory is valid at scales
as small as the coherence length, $\xi$.
Written in terms of dimensionless variables, the
Ginzburg-Landau energy functional is

\begin{equation}
E_{GL} = \int \left[ \begin{array}{c} \left| \psi^{*} \left( 
\vec{\nabla}/i - \vec{A} \right) \psi \right|^{2} - 
\left| \psi \right|^{2} \\ + \frac{\kappa^{2}}{2} \left| \psi
\right|^{4} + B^{2}  
\end{array} \right] d^{3}x.
\label{Eq:eq1}
\end{equation}  
In Eq.~\ref{Eq:eq1}, $\vec{A}$ is the 
vector potential, and the magnetic field 
$\vec{B}(\vec{r}) = \vec{\nabla} \times \vec{A}$.
$\kappa \equiv \frac{\lambda}{\xi}$ is the Ginzburg parameter, the 
ratio of the magnetic penetration depth $\lambda$ 
and the coherence length.  Since the experiments on 
artificial defect arrays typically involve thin
film geometries, we focus on them in   
our treatment.  In considering the thin film 
limit, it is convenient to integrate by parts and replace 
the electromagnetic contribution, $|\vec{\nabla} \times \vec{A}|^2$ 
with the quantity $\vec{J} \cdot \vec{A}$.  As the thin film limit 
is approached, one finds that $\psi$, $\vec{A}$, and $\vec{J}$ 
vary only very slowly in the $\hat{z}$ direction.  The integration
in the $\hat{z}$ direction is well-approximated by multiplication by $d$,
the thickness of the film, and we may write 
\begin{equation}
E_{GL} = \int \left[ \begin{array}{c} \left| \phi^{*} \left(
\vec{\nabla}/i - \vec{A} \right) \phi \right|^{2} -
\left| \phi \right|^{2} \\ + \frac{\lambda_{eff}}{2} \left| \phi
\right|^{4} + \vec{j} \cdot \vec{A} 
\end{array} \right] d^{2}x.
\label{Eq:eq1.5}
\end{equation}
In Eq.~\ref{Eq:eq1.5}, we use the definitions $\phi \equiv d \psi$,
$\vec{j} \equiv d \vec{J}$, and (following standard  
convention~\cite{pearl}) $\lambda_{eff} \equiv \frac{\kappa^{2}}
{d}$. 
 
In this work, we report on results 
obtained for $\lambda_{eff} = 64$ for defects larger than the coherence 
length.  As is typically the case for thin films, the new length scale 
$\lambda_{eff}$ is considerably larger than $\lambda$.   
In fact, this choice of $\lambda_{eff}$ is deep 
enough into the high $\lambda_{eff}$ limit that, apart from scale factors 
in $E_{GL}$ and $\phi$, the results vary little
as $\lambda_{eff}$ is increased.   

To analyze the behavior of the vortex near the pinning center,
we employ a mean field approach in which
one
minimizes $E_{GL}$ for a fixed vortex location to find $\psi$, $\vec{A}$,
and the current $\vec{j}$.  
Our strategy 
for calculating $\psi$ and $\vec{A}$ self-consistently involves first holding 
$\phi$ fixed at some initial guess,
and minimizing $E_{GL}$ with respect
to  $\vec{A}$ and $\vec{B}$.  Next, we fix $\vec{A}$ 
and $\vec{B}$ and minimize with respect to $\phi$.   
These steps are iterated until 
changes in the variables become negligible.  In the case of 
an external current, minor adjustments are needed which 
will be discussed later.  
We implement this self-consistent approach numerically by
dividing the unit cell into a fine lattice of small unit cells.  In this discrete
scheme, $\phi(\vec{r})$ is replaced by $\phi_{ij}$ with $ij$ specifying
a grid point on a square lattice, while $\vec{A}_{ij}$
and $\vec{j}_{ij}$ are defined on nearest-neighbor links
between the grid points.
Derivatives in Eq.~\ref{Eq:eq1} are replaced by the 
corresponding finite differences.  
To model the defect array as accurately as 
possible, one desires a fine grid; we find
that with a $128 \times 128$ mesh our results are well-converged
with respect to the discretization.  

To see how one minimizes $E_{GL}$ under the constraint of a
specified vortex location, it is 
useful to write the current $\vec{j}$ in  terms of the order parameter $\phi$ 
and the vector potential $\vec{A}$.  By minimizing $E_{GL}$  with respect
to the vector potential and employing a Maxwell equation one has 
\begin{eqnarray}
\label{Eq:eq2}
\vec{j} &=& \frac{1}{2} \left[ \phi^{*} \left( \vec{\nabla}_{\parallel}/i - 
\vec{A} \right) \phi + \phi \left( -\vec{\nabla}_{\parallel}/i - \vec{A} \right) \phi^{*} 
\right] \\
&=& \left| \psi \right|^{2} \left( \vec{\nabla}_{\parallel} \theta - 
\vec{A} \right) . \nonumber
\end{eqnarray} 
In Eq.~\ref{Eq:eq2} we have used for the order parameter $\phi = \left| \phi 
\right | e^{i\theta}$, and the gradient $\vec{\nabla}_{\parallel} = (\partial_{x},
\partial_{y}, 0)$. 
The familiar fluxoid quantization condition\cite{tinkham} arises from 
the requirement that the order parameter be single valued, i.e. $\oint
\vec{\nabla} \theta \cdot
d \vec{s} = 2 \pi n_{v}$ with $n_{v}$ an integer.  Hence, 
in terms of $\vec{j}$ and $\vec{A}$,  
\begin{eqnarray}
\label{Eq:eq3}
2 \pi n_{v}(ij) &=& \oint \left( \vec{J}/\left| \phi \right|^{2} \right) \cdot d\vec{s} + 
\oint \vec{A} \cdot d\vec{s}\\  
&=& \oint \left(  \vec{j}/\left| \phi \right|^{2} \right)\cdot d\vec{s} \nonumber
+ \Phi_{B} 
\end{eqnarray}
In the second part of Eq.~\ref{Eq:eq3}, $\Phi_{B}$ is the total magnetic 
flux passing through the area of the contour, which we conveniently 
choose to be the small unit cell associated with the grid point $ij$, 
while $n_{v}(ij)$ is the total number of ``fluxoid quanta'' 
contained in the contour of integration\cite{tinkham}. 
It is through $n_{v}$ that the vortex location(s)
in the full unit cell of the system may be fixed:
$n_{v}=0$ except at the grid points where we
wish to place a vortex, for which $n_{v}=1$.  As discussed in the 
Appendix, one can easily calculate $\Phi_{B}$ and $\vec{A}$ self-consistently 
in the thin film limit from the current $\vec{j}$.  
Hence, armed  with knowledge of $|\phi|$ and some specified realization
of $n_{v}(ij)$, one can
solve for $\vec{j}$ and $\vec{A}$ via Eq.~\ref{Eq:eq3}.  Using the expression 
for the current given in Eq.~\ref{Eq:eq2}, 
one obtains $\vec{\nabla} \theta$; inserting $\vec{\nabla} \theta$ and $\vec{A}$ 
into Eq.~\ref{Eq:eq1} yields an expression depending only on $\left| \phi \right|$
and $\lambda_{eff}$ which we minimize with respect to $\left| \phi \right|$ to obtain
 the order parameter modulus.  

Our calculational method is easily generalized to include a 
supercurrent $\vec{j_{o}}$ flowing across the system, allowing the 
effects of a depinning current to be probed.  This is accomplished by 
introducing
a constant planar vector potential $\vec{a}_{o}$.  $\vec{a}_{o}$ causes a shift
in the current given by $\Delta \vec{J} = -|\phi|^{2} \vec{a}_{o}$.  

In the above method, we have been careful to explicitly take into
account the thin film geometry typical of experiments on nanoscale
defect arrays.  However, we note that this approach may easily be 
modified to handle the bulk case, 
in which the vortices resemble filaments of 
magnetic flux instead of pancake structures. 
We have performed a few such calculations, 
and find that for antidot systems the results obtained
are quite similar to the large $\lambda_{eff}$ results we report here.
This may be understood in terms of the effective magnetic penetration
depth for a thin film, $\lambda_{eff}
= \frac{\lambda^{2}}{d}$, which is typically much larger 
than the bulk value $\lambda$\cite{pearl}. 
This means that the energy stored in the magnetic field generated
by the supercurrents is quite small, so that the fact that the field 
varies as one moves out of the plane has little impact 
on the state of the system.  The resulting energy functional is thus 
nearly identical to the bulk three dimensional case in the large
$\kappa$ limit, with columnar antidots and 
vortices.

\section{Results}
\subsection{Zero Current}

\begin{figure}
\begin{center}
\centerline{\psfig{figure=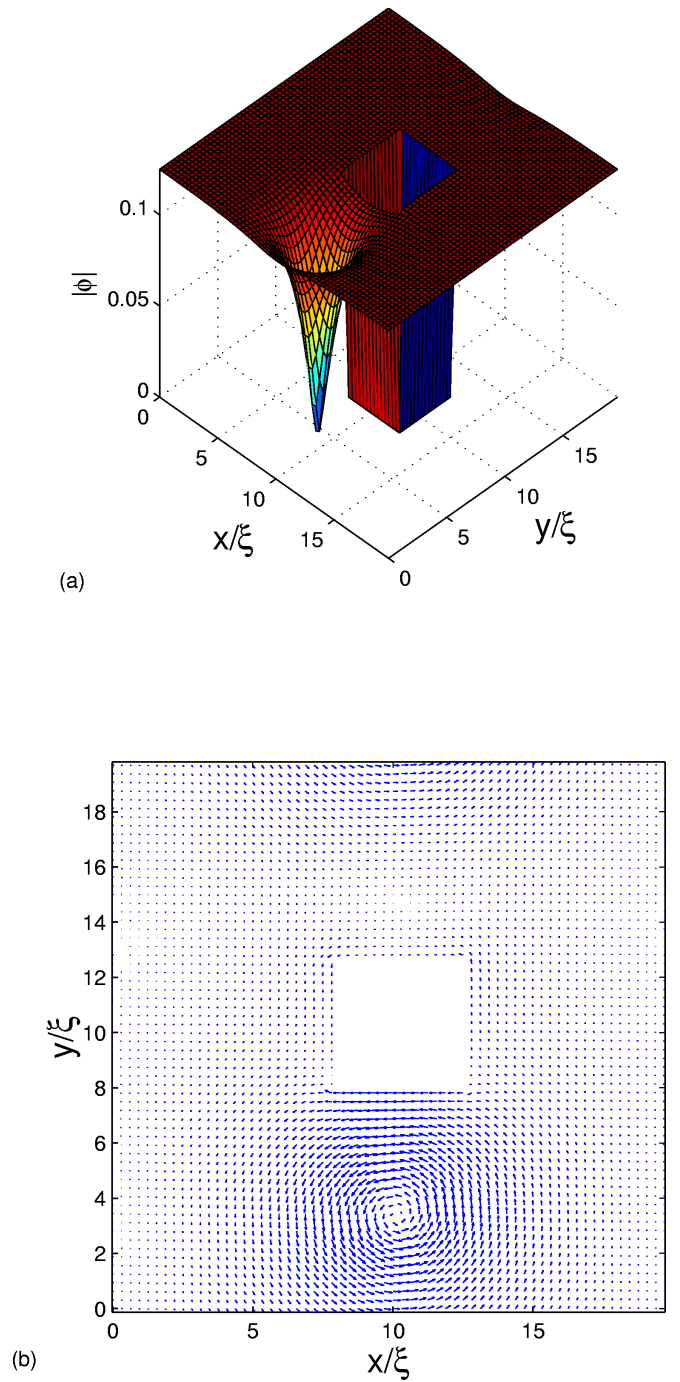,width=3.4in}}
\vspace{5mm}
\caption{A perspective plot of the order parameter modulus, $|\phi|$ (a) and current 
image (b) in a periodic antidot array just prior to the transition.  
Currents are localized about the vortex
core, which has a compact structure.} 
\label{Fig:fig2}
\end{center}
\end{figure}

\begin{figure} 
\begin{center}
\centerline{\psfig{figure=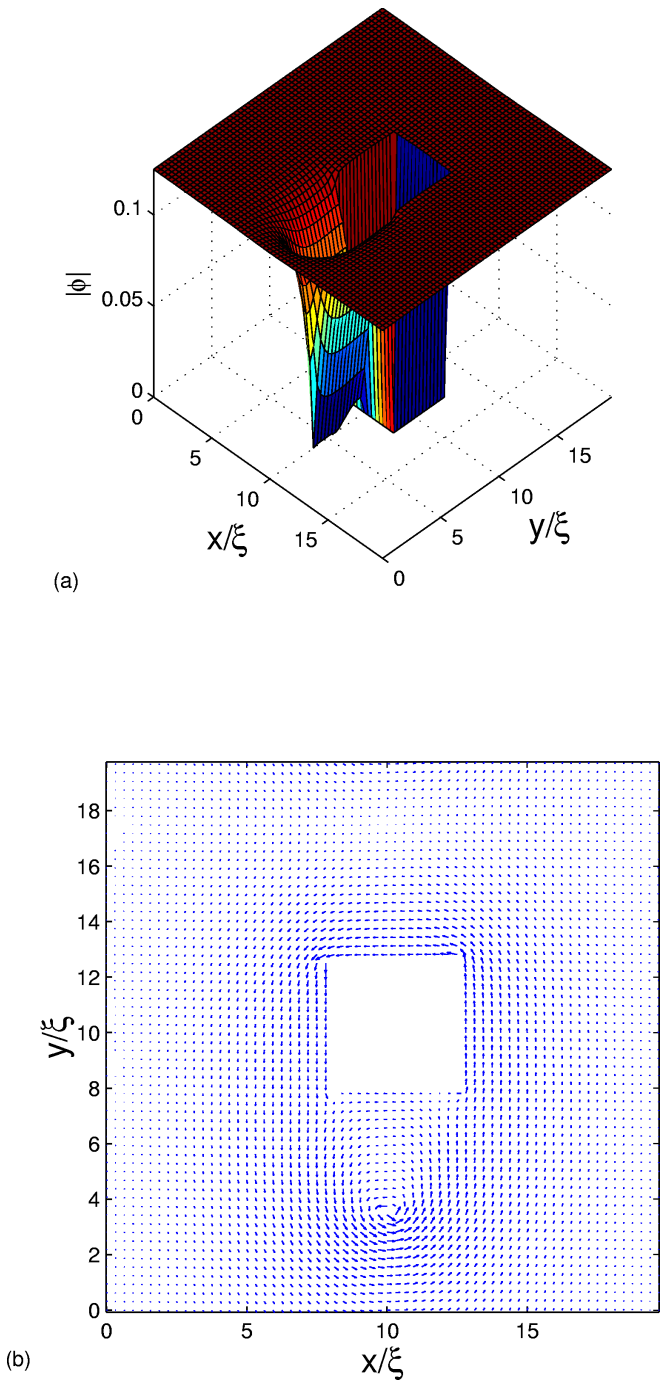,width=3.4in}}
\vspace{5mm}
\caption{Order parameter perspective plot (a) and currents (b) in the periodic 
array just after the transition.  
The currents circulate about the vortex-defect pair, and the vortex core has 
a string extending all the way to the defect edge.}
\label{Fig:fig3}
\end{center}
\end{figure}

At large vortex-defect separations, the core has 
the usual compact structure with supercurrents 
localized about it.  Fig.~\ref{Fig:fig2}
presents a perspective plot of $\left| \phi \right|$, as well as a vector plot of 
the currents.  The distances shown  
are in units of the coherence length, $\xi$.  
Our choice of a unit cell with side spanning 20 coherence lengths  
and an antidot $5 \xi$ in size
is typical of many of the nanoscale arrays studied experimentally.
As the flux quantum nears the defect 
edge, it eventually reaches a critical distance $d_{c}$ 
(typically several $\xi$, with precise value 
depending on the dot size and shape), where  
there is a sudden dramatic change.  Fig.~\ref{Fig:fig3} illustrates
the situation after the transition:
the vortex core has developed
a {\it string} extending from 
the flux quantum position to the defect edge; simultaneously the current now 
encircles the vortex-defect pair, and the magnetic flux created by these
currents spreads over a larger area. 
The string 
is energetically favorable because it allows the formerly 
dense current of the vortex to spread out (engulfing the 
defect in the process), thereby reducing the kinetic 
energy of the state. 
 

\begin{figure}
\begin{center}
\centerline{\psfig{figure=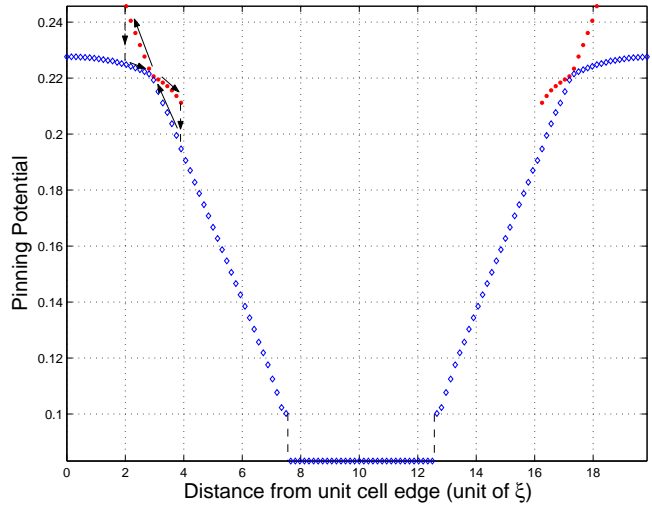,width=3.4in}}
\vspace{5mm}
\caption{Plot of the pinning potential showing broad linear 
regions and a gap in the potential at the defect boundary.  The 
horizontal axis measures the distance of the vortex from the 
edge of the unit cell.  Arrows indicate 
hysteresis in the calculation, a hallmark of a first order
transition.} 
\label{Fig:fig4}
\end{center}
\end{figure}

When a net supercurrent flows across a superconductor,
a Lorentz force is exerted on vortices; 
if these move in response energy
is dissipated.  In an artificial defect array, the 
driving force can be balanced by
a pinning force given by the gradient of $E_{GL}$
with respect to the vortex position.  Thus, one may 
regard $E_{GL}$ as the pinning potential. 
Fig.~\ref{Fig:fig4} shows this 
as a function of distance from the unit cell edge,
which can be seen to have
three distinct regions.  For large separations, where 
the vortex core is compact, 
the pinning potential decreases relatively slowly.  
As the separation decreases it eventually crosses 
$d_{c}$ and
the core-string structure appears.  Hysteresis in the calculations, 
indicated with arrows, suggests that 
the transition is first 
order.  However, as will be seen below, a finite current flowing across the unit 
cell changes the character of the transition, rendering
it continuous.  Below the transition one observes 
nearly linear  behavior of the pinning potential,  
suggesting that the
string carries an energy proportional to its length. 
The third region is  announced by a discontinuous
jump as the vortex is absorbed by 
the defect, followed by a perfectly flat region inside
the defect.  This  abrupt jump is due 
to a sudden transformation of the order parameter.  Immediately 
prior to absorption, constant Cooper pair density contours 
near the flux quantum have a semicircular profile.
As the vortex enters the defect, this ``semi-vortex'' vanishes,
removing a finite amount of energy
for an infinitesimal change in the vortex position.  In later
discussion, the $\Delta_{ej}$ shall be used to refer to this 
jump in energy.

\subsection{Finite Current}

When a net current flows across the superconducting  
film, a Lorentz force 
is exerted on the current distribution by the 
external magnetic field.  
This nonconservative force, given by $\frac{1}{c} \mathbf{\Phi_{o}}\times
\mathbf{J}$~\cite{tinkham}, acts in opposition to $\vec{\nabla}E_{GL}$.
In terms of our units, the magnitude of the Lorentz force is $4 \pi j^{'}$,
where $j^{'}$ is the net current magnitude (In CGS units, $J = 
\frac{\hbar c^{2} }{8 \pi d e \xi^{3}}j^{'}$).  $E_{GL}$ also depends on 
$j^{'}$.  As will be seen, this dependence is particularly manifest 
near the depinning threshold.   

\begin{figure}
\begin{center}
\centerline{\psfig{figure=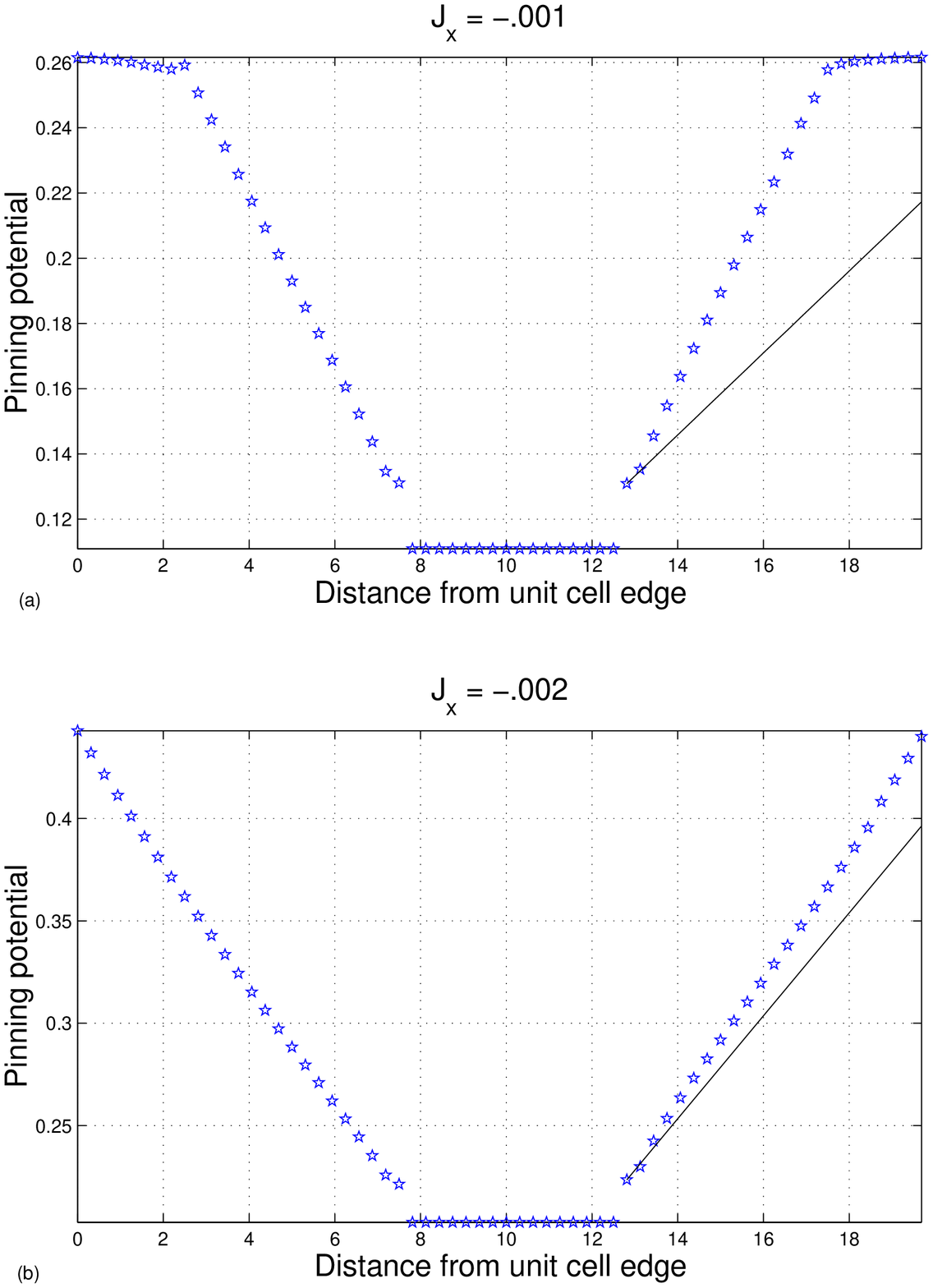,width=3.2in}}
\vspace{5mm}
\caption{Pinning potentials for $J_{x} = -.001$ (a) and $J_{x} = -.002$ (b) 
The exit barrier for the flux quantum does not diminish despite increasing
drive current.  The slope of the dark line in both plots is the Lorentz 
force.} 
\label{Fig:fig5}
\end{center}
\end{figure}

Figure~\ref{Fig:fig5} indicates the evolution of $E_{GL}$ as the 
current is increased.  As the two graphs in Fig.~\ref{Fig:fig5}
reveal, even for currents a reasonable fraction (17\% and 33\%
respectively) of the 
bulk critical current $J_{c}$, $E_{GL}$ is 
similar to that of the zero-current case, except that 
the breadth of the linear region increases with drive current.
Ultimately, as can be seen in the potential curve corresponding
to the larger current, the linear behavior extends to either 
edge of the unit cell.  This suggests there is a range of net current
in which the vortex core always exists as a stringlike structure.  
Another important feature is the fact that the jump in energy when 
the fluxoid exits the dot (which we call the ``ejection barrier''
$\Delta_{ej}$)
does not appreciably diminish as the drive current $j^{'}$ is increased.  
An interesting observation, apparent in Figs.~\ref{Fig:fig5},
is that the Lorentz force increases with increasing current
while the slope of the conservative part of the pinning 
potential remains nearly constant.  Eventually, the 
Lorentz force exceeds the slope of $E_{GL}$ vs. fluxoid
position before the jump $\Delta_{ej}$ has been eliminated.
This suggests that the ejection of the vortex with increasing
drive current will be sudden at low temperatures, i.e., once the flux quantum is
able to exit the dot, the string tension will not be sufficiently
strong to keep the vortex bound.  We will see below that this
is indeed the case.  At higher temperatures, when $T > \Delta_{ej}$, 
one expects to see depinning for somewhat lower drive currents.

\begin{figure}
\begin{center}
\centerline{\psfig{figure=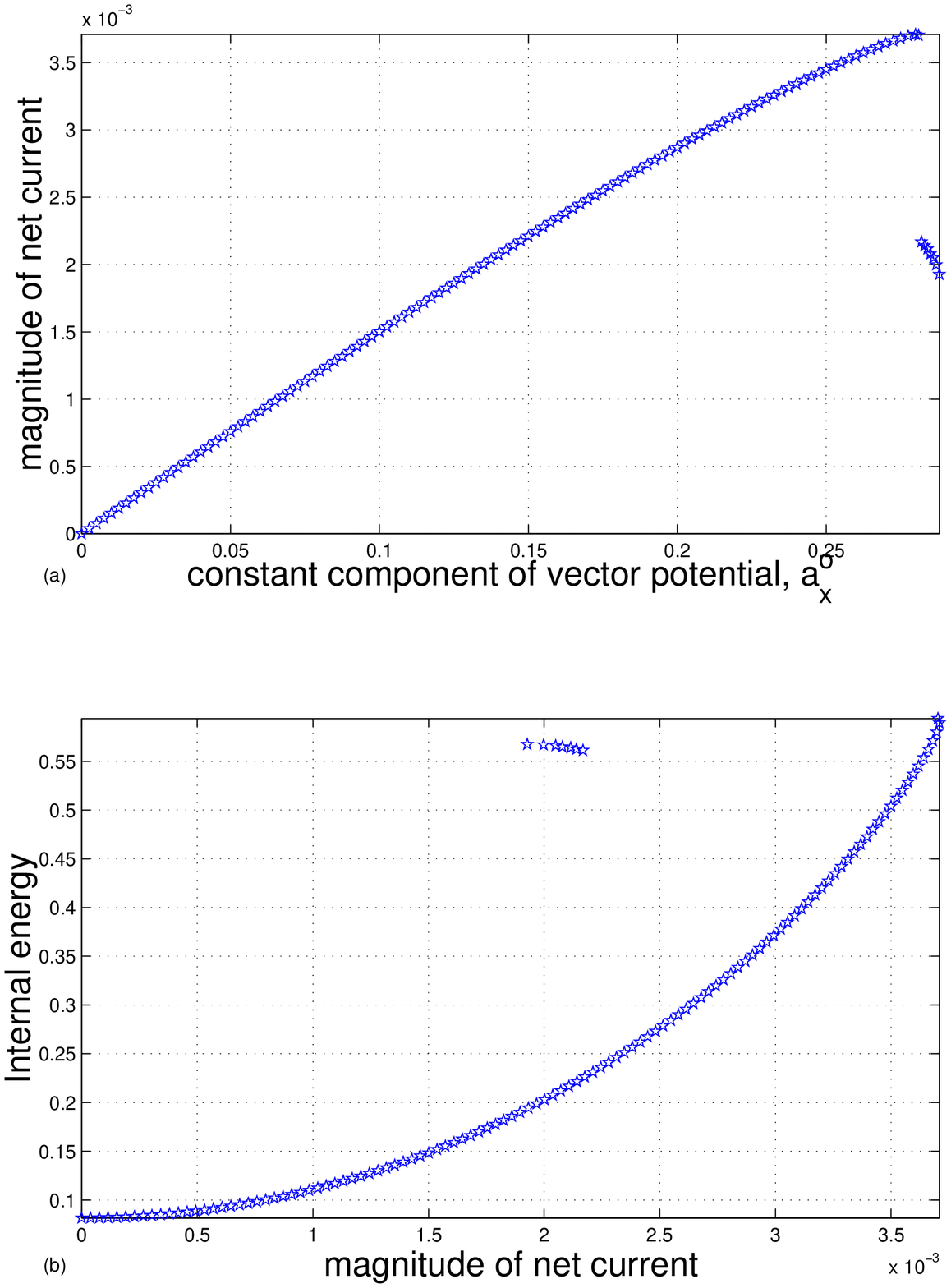,width=3.4in}}
\caption{Plot of net current as a function of $a_{x}^{0}$ (top)
and Ginzburg-Landau energy versus the current (bottom)}
\label{Fig:fig4.5}
\end{center}
\end{figure}

\begin{figure}
\begin{center}
\centerline{\psfig{figure=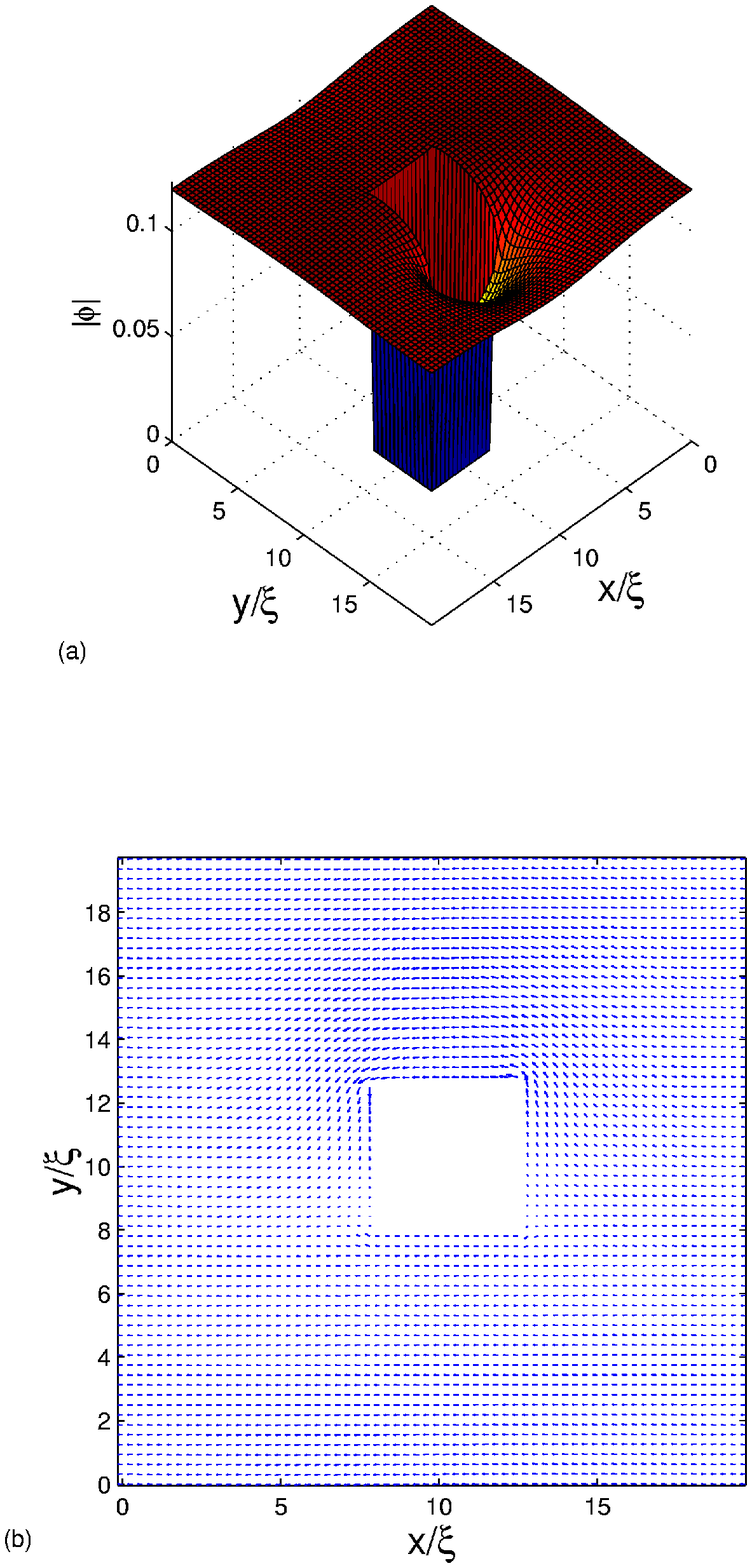,width=3.4in}}
\vspace{5mm}
\caption{A perspective plot of the order parameter modulus, $|\phi|$ (a) and
current
image (b) in a periodic antidot array just prior to the transition.
A core string has not appeared.}
\label{Fig:fig4.7}
\end{center}
\end{figure}

\begin{figure}
\begin{center}
\centerline{\psfig{figure=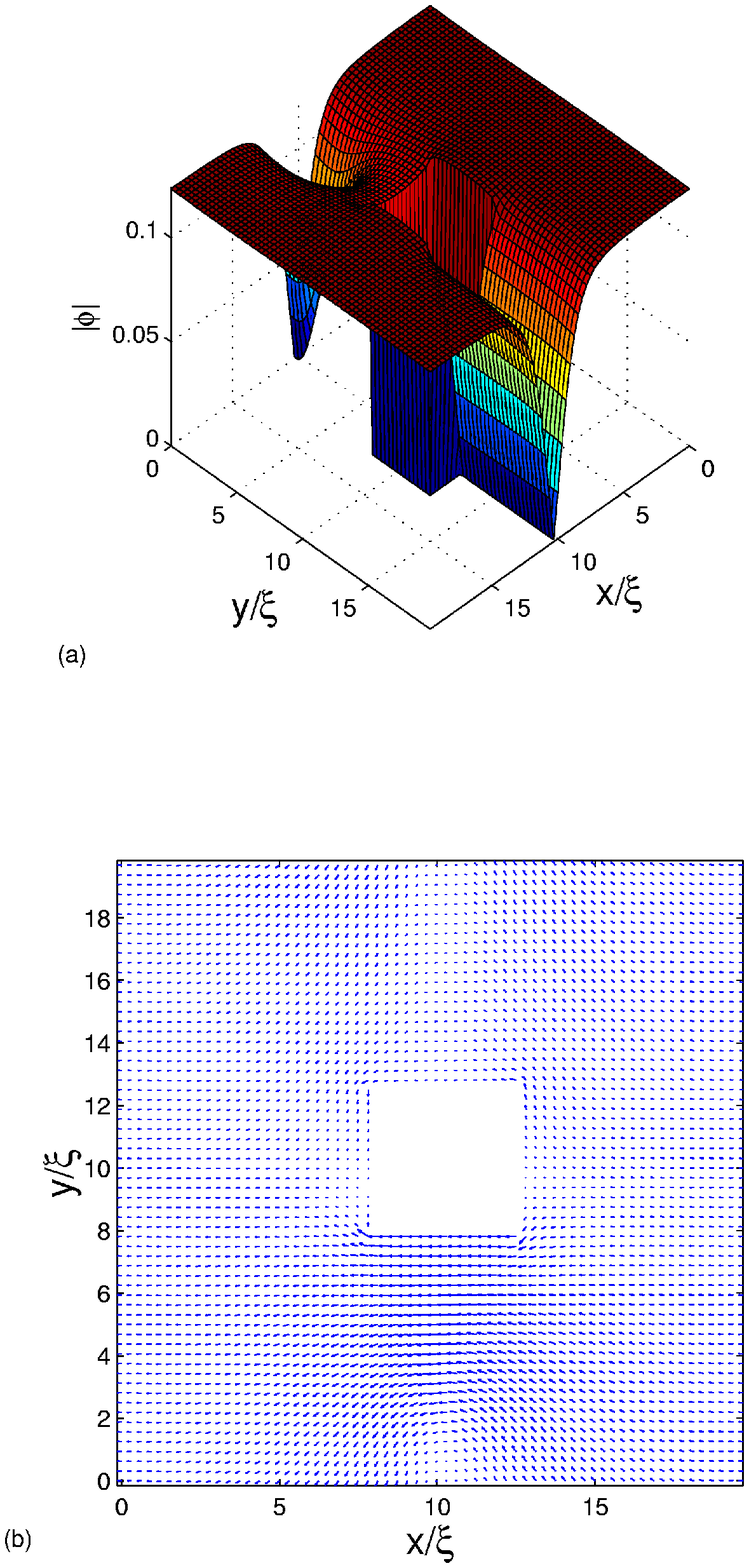,width=3.4in}}
\vspace{5mm}
\caption{A perspective plot of the order parameter modulus, $|\phi|$ (a) and
current
image (b) in a periodic antidot array with core string, just
beyond depinning} 
\label{Fig:fig4.8}
\end{center}
\end{figure}

Depinning occurs when the net current is made large
enough to eliminate 
$\Delta_{ej}$.  This occurs suddenly as one increases the current 
beyond a certain threshold, $j^{'}_{dp}$.  As Figure~\ref{Fig:fig4.5} 
indicates, the net supercurrent drops abruptly; presumably the 
remainder of the drive current appears in resistive channels.  The
fact that the solutions to the Ginzburg-Landau theory have a 
maximum possible value $j^{'}$ even for large $a_{o}^{x}$ is highly
analogous to what happens in thin superconducting wires and
films~\cite{tinkham}.
As in those cases, this maximum current should be identified as 
a critical current; in our case, it is the depinning 
current $j^{'}_{dp}$. 
As illustrated in Figure~\ref{Fig:fig4.7} and Figure~\ref{Fig:fig4.8},
the depinning transition is marked by the sudden emergence of a
core string.  The string, a normal region which blocks supercurrents,
accounts for the sudden drop in the supercurrent flow across the 
unit cell.
For the system parameters we have chosen, 
the depinning current, $j^{'}_{dp}$ is $62 \%$ of the 
bulk critical current $j^{'}_{c}$.  

\begin{figure}
\begin{center}
\centerline{\psfig{figure=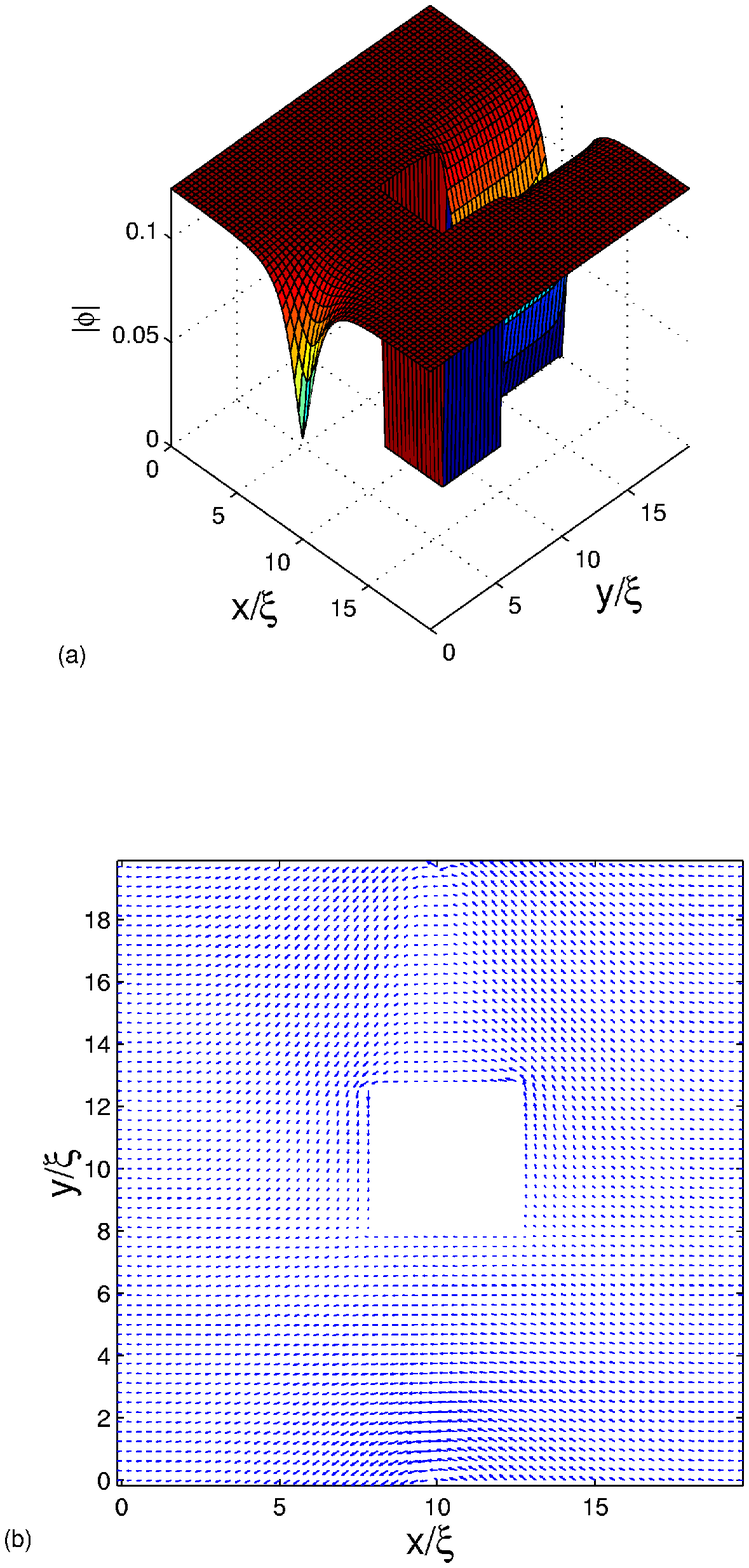,width=3.4in}}
\vspace{5mm}
\caption{A perspective plot of the order parameter modulus, $|\phi|$ (a) and
current
image (b) in a periodic antidot array just prior to the transition.
The core string is directed away from the pinning defect}  
\label{Fig:fig6}
\end{center}
\end{figure}

\begin{figure}
\begin{center}
\centerline{\psfig{figure=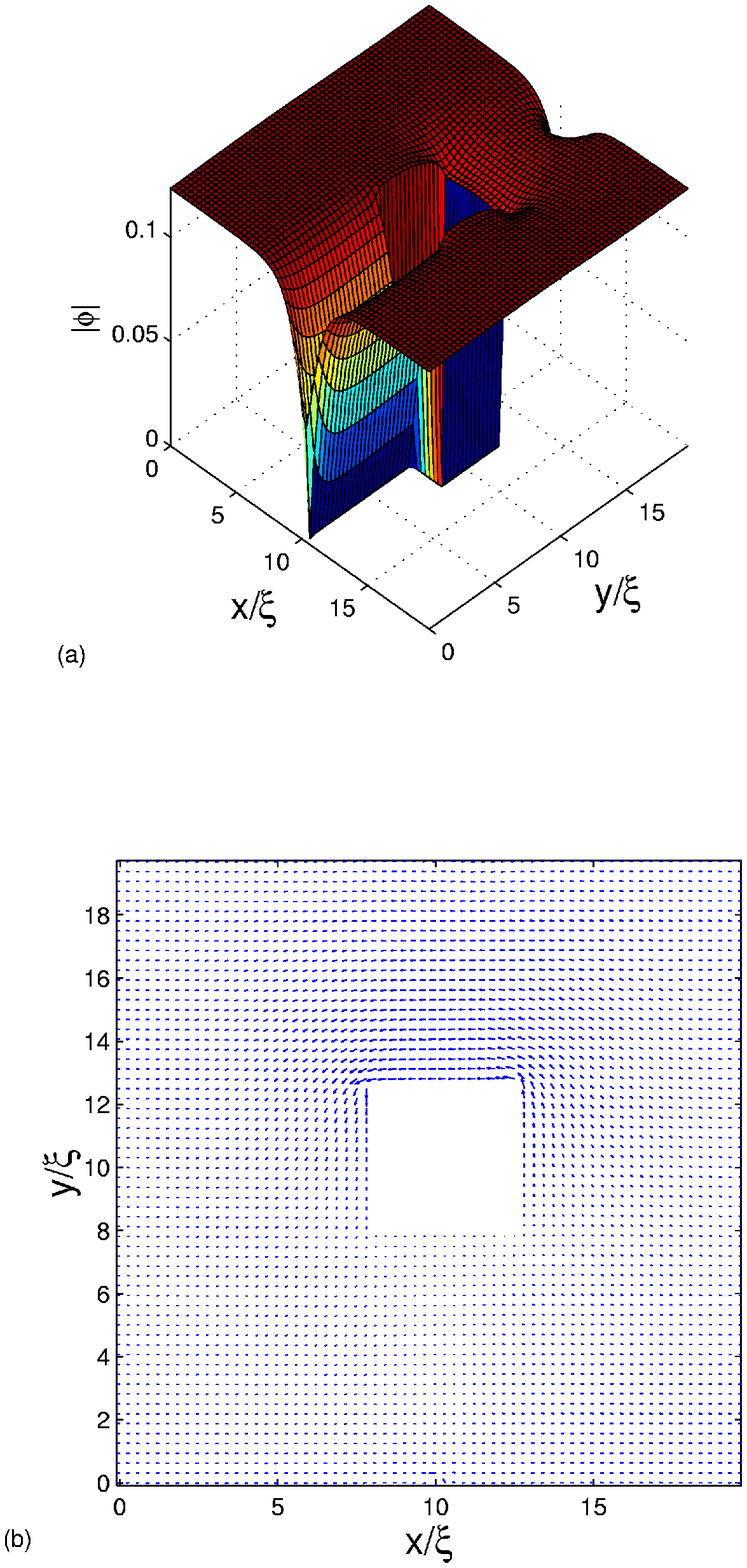,width=3.4in}}
\vspace{5mm}
\caption{A perspective plot of the order parameter modulus, $|\phi|$ (a) and
current
image (b) in a periodic antidot array just after the ``unwrapping'' transition.
The core string is attached to the adjacent ``destination'' defect.  Currents have 
moved from the pinning defect and encircle the ``destination'' defect.}
\label{Fig:fig7}
\end{center}
\end{figure}

Strikingly, currents capable of depinning the 
vortex do so without forming 
a flux quantum that has a compact core.  In fact, as the 
vortex moves across the unit cell, its core always has 
a stringlike structure.  As the flux quantum is 
driven from the pinning defect, a core ``string'' forms
connecting the vortex to the edge of the pinning defect.  
Even before the flux quantum emerges from the pinning defect,
this string is already partly formed. As the
vortex traverses the unit cell toward the neighboring 
defect, the core string lengthens.  This continues until
the flux quantum reaches the unit cell edge, where  
the magnetic flux and currents
unwrap from the pinning defect and engulf the 
adjacent defect.  Figure~\ref{Fig:fig6} and Figure~\ref{Fig:fig7}
illustrate this transition.  Figure~\ref{Fig:fig6} displays the 
order parameter modulus and current patterns just prior to the 
transition, while Figure~\ref{Fig:fig7} portrays the system 
just after the currents and flux have surrounded the neighboring 
defect.

     We conclude this section by noting that the depinning process
found for typical antidot arrays is markedly different than 
previous expectations, which commonly assume that in depinning
the vortex takes a form similar to Figure~\ref{Fig:fig2}. 
We expect the latter scenario to hold when the interdot 
separation is very large compared to the coherence 
length $\xi$.  The different forms of depinning might be 
detectable in voltage noise experiments, which are analogous
to shot noise experiments for electrons in conductors.  When 
the vortices are small and compact, it is natural to suppose 
they will act as particles and, when depinned, yield a 
broadband noise proportional to the voltage across the 
system.  If one increases the temperature and corresponding 
coherence length, the string depinning behavior described 
above must eventually apply.  This eliminates short
wavelength and presumably high frequency dynamics from 
the system.  We thus expect a suppression of high frequency 
noise to accompany the transition into string-like depinning.

\section{Experimental implications} 
Beyond the voltage noise behavior described above,
   the profile of the pinning potential shown in Fig.~\ref{Fig:fig4} suggests
a set of measurements one might perform to seek an experimental signature of 
the unique pinning phenomena discussed above.  Past the depinning 
threshold, core strings moving among the antidots form linear 
channels which might appear as an anisotropy signal in neutron scattering measurements.
Another possible test would involve working at temperatures 
large compared to $\Delta_{ej}$, but small compared to the 
energy at the unwrapping transition, as in Fig.~\ref{Fig:fig3}.  In this 
situation, an AC driving
force on the vortices, of magnitude small enough to leave them
pinned by the strings (as might be provided
in a vibrating reed experiment\cite{lance2})  
would allow the
vortices to ``rattle around'' in the linear part of the pinning well
illustrated in Fig.~\ref{Fig:fig4} and produce losses,
whereas a DC force of equal magnitude would be
dissipationless.  Upon lowering of the temperature
the vortices would be captured by the defects, leading
to lossless supercurrents for both AC and DC driving
forces.  An observation of these effects would yield
indirect confirmation of the form of 
the pinning potential we find.  Something 
like this may recently have been observed\cite{lance2}. 

Finally, a consequence of our results is that the depinning 
current should increase sharply as $T$ is lowered below
$\Delta_{ej}$.  For $T > \Delta_{ej}$, vortices will be 
thermally excited above the ejection barrier, and depinning 
can occur when the slope of $E_{GL}$ versus fluxoid position 
is equal to the Lorentz force.  For $T \ll \Delta_{ej}$ the 
current must be large enough to eliminate the ejection barrier 
to depin the vortex~\cite{caveat}. 

Another interesting possibility is that a unique thermal depinning 
may occur as the temperature is increased in 
the regime of linear pinning.  The presence of 
a string 
suggests that this may carry entropy at finite
temperature, much as is the case of polymers.
This entropy is proportional to the string length
and temperature, and at high enough temperatures
may overwhelm the energy
per unit length found in our mean-field
calculations.  In analogy with polymer behavior \cite{mrstring},
this leads to unbounded growth of the string and
effective depinning of the vortex.
However, it is not clear whether the string remains sufficiently
well-defined at the temperatures necessary for
proliferation that the polymer analogy remains valid up to the transition.
Further research into this possibility is currently underway.  

\section{Summary}

     Using Ginzburg-Landau theory, we have given a detailed 
treatment of the microscopic aspects of pinning phenomena in nanoscale periodic 
arrays.  Strikingly, we see an apparent first order 
transition involving the creation of a string connecting the vortices to
the defect,
and an accompanying abrupt transformation of the supercurrent
and the magnetic field it generates. 
The string configuration leads 
to a region of linear pinning.
Absorption of the vortex by the antidot is marked by a jump
in the pinning potential, $\Delta_{ej}$.  Depinning occurs abruptly when 
the drive current exceeds $J_{dp}$.  
Various aspects of the pinning
potential should be observable
in experiment.

{\it Acknowledgments--}
The authors would like to thank L. E. DeLong, S. B. Field,
and J. B. Ketterson for useful discussions.
This work was supported by NSF Grant No.
and DMR-0108451.

\appendix

\vspace{3mm}
\section{Calculation of $\Phi_{B}$}
\vspace{3mm}

In this Appendix, we outline the calculation of the   
magnetic flux density in terms of the supercurrents  
in the thin film.  
The lattice variables mentioned in this work
are discrete approximations of continuum 
quantities.  For the sake of brevity, however,
the analysis in this Appendix is given 
in the continuum limit.  The calculation 
sketched here is largely parallel to an earlier 
work in the context of the London theory~\cite{pearl}. 

    We adopt units for which the current $\vec{J^{'}}$ 
satisfies $\vec{J^{'}} = \vec{\nabla} \times \vec{B}$.
Taking advantage of the fact that $\vec{\nabla} \cdot
\vec{A} = 0$ in the London gauge, we find that
 
\begin{equation}
\vec{J^{'}} = -\nabla^{2} \vec{A}
\label{Eq:eqap1}
\end{equation}
Fourier transforming Eq.~\ref{Eq:eqap1} yields

\begin{equation}
\vec{J^{'}_{\mathbf{k}}} = -k^{2} \vec{A_{\mathbf{k}}}
\label{Eq:eqap2}
\end{equation}
Defining $k_{\parallel}^{2} \equiv (k_{x}^{2} 
+ k_{y}^{2})$, we have 
\begin{equation}
 J^{'x,y}_{\mathbf{k}} = 
-\left( k_{z}^{2} + k_{\parallel}^{2} \right)  
A_{\mathbf{k}}^{x,y} 
\label{Eq:eqap3}
\end{equation}
To express $\vec{J^{'}_{\mathbf{k}}}$ in terms of the 
two dimensional currents flowing in the plane of the 
film one uses the fact that $\vec{J^{'}}$ is essentially
constant in the $\hat{z}$ direction in the thin film
limit.  Hence, we have 
\begin{equation}
J^{'x,y} =  \left \{ \begin{array}{c} 
J^{'x,y}_{2d}, - \frac{d}{2} \leq
z \leq \frac{d}{2};\\ 0 $~elsewhere$ \end{array} \right \}
\label{Eq:eqap4}
\end{equation}
As the film thickness diminishes to zero, we
obtain
\begin{equation}
J^{'x,y}|_{d \rightarrow 0} = 
  \delta \left( z \right)  j^{'x,y}_{2d}
\label{Eq:eqap5}
\end{equation}
Fourier transforms equation~\ref{Eq:eqap5}
yields
\begin{equation}
J^{'x,y}_{\mathbf{k}} =  
\frac{1}{L_{z}}  j^{'x,y}_{\mathbf{k_{\parallel}}}
\label{Eq:eqap6}
\end{equation}
In Eq.~\ref{Eq:eqap6} above, $\mathbf{k_{\parallel}}$ denotes the 
plane wave vector.  We infer from Eqs.~\ref{Eq:eqap3},~\ref{Eq:eqap6}
that 
\begin{equation}
 A^{x,y}_{\mathbf{k}} = -\frac{1}{L_{z} 
\left( k_{z}^2 + k_{\parallel}^{2} \right)}  
j^{'x,y}_{\mathbf{k_{\parallel}}}
\label{Eq:eqap7} 
\end{equation}
To obtain $A^{x,y}|_{z=0}$, and $B^{z}|_{z=0}$, an
inverse Fourier transformation of Eq.~\ref{Eq:eqap7} is needed.  
One finds that 
\begin{eqnarray}
 A^{x,y}|_{z=0} &=& \int \!\!\int L_{x}L_{y} \int 
 \frac{- (2 \pi)^{-3}  j^{'x,y}_{\mathbf{k_{\parallel}}}}
 { \left( k_{z}^{2} + k_{\parallel}^{2} \right) }
 d^{3}k\\
\label{Eq:eqap8}
 &=& \frac{L_{x}L_{y}}{2 \left( 2 \pi \right)^{2}} \int \!\! \int \frac{
 -  j^{'x,y}_{\mathbf{k_{\parallel}}}} 
 {k_{\parallel}} dk_{x} dk_{y} 
\end{eqnarray} 
To obtain $B^{z}|_{z=0}$, we use the fact that
\begin{equation}
B^{z}|_{z=0} = - \frac{\partial}{\partial y}A_{x}|_{z=0} + 
\frac{\partial}{\partial x}A_{y}|_{z=0} 
\label{Eq:eqap9}
\end{equation}
In Fourier space, one has 
\begin{equation}
B^{z}_{\mathbf{k_{\parallel}}}|_{z=0} = 
-k_{y}A^{x}_{\mathbf{k_{\parallel}}}|_{z=0}
+ k_{x}A^{y}_{\mathbf{k_{\parallel}}}|_{z=0}
\label{Eq:eqap10}
\end{equation}
Hence, we find that 
\begin{equation}
B^{z}|_{z=0} = \frac{L_{x}L_{y}}{2(2\pi)^{2}} \int \!\! \int \frac{\left(
k_{y}j^{'x}_{\mathbf{k_{\parallel}}} - 
k_{x}j^{'y}_{\mathbf{k_{\parallel}}} 
\right)}{k_{\parallel}} dk_{x} dk_{y}.
\label{Eq:eqap11}
\end{equation}
To calculate $\Phi_{B}^{ij}$, the magnetic flux passing 
through the small unit cell at the point $ij$, we 
integrate $B^{z}|_{z=0}$ over the area of the small square.
In the discretization scheme employed in this work, we 
use the approximation
\begin{equation}
\Phi_{B}^{ij} = a^{2}B^{z}_{ij}|_{z=0},
\end{equation}
where $a^{2}$ is the area of the small unit cell.


\begin{thebibliography}{99}
\bibitem{caveat}
However, it should be noted that the change in depinning 
current is of order $50 \%$, so that this phenomenon may
be difficult to sort out from other temperature 
dependence entering through the coherence length and 
other GL parameters.
\end{thebibliography}
\end{document}